\documentclass[aps,amsfonts,prl,showpacs,nobibnotes,nofootinbib,%
tightenlines,twocolumn]{revtex4}

\newcommand{\vek}[1]{\mathbf{#1}}

\begin{document}

\title{Casimir Energies in Light of Quantum Field Theory}

\author{N.~Graham}
\affiliation{Department of Physics and Astronomy,
University of California at Los Angeles, Los Angeles, CA  90095}

\author{R.~L.~Jaffe}
\author{V.~Khemani}
\author{M.~Quandt}
\author{M.~Scandurra}
\affiliation{Center for Theoretical Physics, Laboratory for Nuclear
Science and Department of Physics, Massachusetts Institute of Technology,
Cambridge, Massachusetts 02139}

\author{H.~Weigel}
\affiliation{Institute for Theoretical Physics,
T\"ubingen University Auf der Morgenstelle 14, D--72076 T\"ubingen, Germany}

\begin{abstract}\noindent We study the Casimir problem as the limit of
a conventional quantum field theory coupled to a smooth background. 
The Casimir energy diverges in the limit that the background forces
the field to vanish on a surface.  We show that this divergence cannot
be absorbed into a renormalization of the parameters of the theory. 
As a result, the Casimir energy of the surface and other quantities
like the surface tension, which are obtained by deforming the surface,
cannot be defined independently of the details of the coupling between
the field and the matter on the surface.  In contrast, the energy
density away from the surface and the force between rigid surfaces are
finite and independent of these complications.
\end{abstract}
\vspace*{-\bigskipamount}

\preprint{UCLA/02/TEP/20, MIT-CTP-3292, UNITU-HEP-10/2002}
\maketitle

The vacuum energy of fluctuating quantum fields that are subject to
boundary conditions has been studied intensely over the half-century
since Casimir predicted a force between grounded metal
plates\cite{casimir,MT,bmm}.  The plates change the zero-point energies
of fluctuating fields and thereby give rise to forces between the
rigid bodies or stresses on isolated surfaces.  The Casimir force
between grounded metal plates has now been measured quite accurately
and agrees with his prediction~\cite{expt1,expt2,expt3}.

Casimir forces arise from interactions between the fluctuating fields
and matter.  Nevertheless, it is traditional to study idealized
``Casimir problems'' where the physical interactions are replaced {\it
ab initio} by boundary conditions.  In this Letter we study under what
circumstances this replacement is justified.  A real material cannot
constrain modes of the field with wavelengths much smaller than the
typical length scale of its interactions.  In contrast, a boundary
condition constrains all modes.  The sum over zero point energies is
highly divergent in the ultraviolet and these divergences depend on
the boundary conditions.  Subtraction of the vacuum energy in the
absence of boundaries only removes the worst divergence (quartic in
three space dimensions).   

The fact that the energy of a fluctuating field diverges when a
boundary condition is imposed has been known for many years\cite{C&D,
Candelas}.  Schemes have been proposed to cancel these divergences by
introducing new, {\it ad hoc\/} surface dependent
counterterms\cite{Symanzik} or quantum boundary functions\cite{Actor}. 
We are not interested in such a formal solution to the problem. The
method of renormalization in continuum quantum field theory
without boundaries (QFT) provides the only physical way to
regulate, discuss, and eventually remove divergences.  Therefore we
propose to embed the Casimir calculation in QFT and study its
renormalization.  After renormalization, if any quantity is still
infinite in the presence of the boundary condition, it will depend in
detail on the properties of the material that provides the physical
ultraviolet cutoff and will not exist in the idealized Casimir
problem.  Similar subtleties were addressed in the context of
dispersive media in Ref.~\cite{Barton}.

It is straightforward to write down a QFT describing the interaction
of the fluctuating field $\phi$ with a static background field
$\sigma(\vek x)$ and to choose a limit involving the shape of
$\sigma(\vek x)$ and the coupling strength between $\phi$ and $\sigma$
that produces the desired boundary conditions on specified surfaces. 
We have developed the formalism required to compute the resulting
vacuum energy in Ref.~\cite{dens}.   Here we focus on Dirichlet
boundary conditions on a scalar field.  Our methods can be generalized
to the physically interesting case of conducting boundary conditions
on a gauge field.

Ideally, we seek a Casimir energy that reflects only the effects
of the boundary conditions and not on any other features of
$\sigma(\vek x)$.  Therefore we do not specify any action for $\sigma$
except for the standard counterterms induced by the $\phi$-$\sigma$
interaction.  The coefficients of the counterterms are fixed by
renormalization conditions applied to perturbative Green's functions. 
Once the renormalization conditions have been fixed by the
definitions of the physical parameters of the theory, there is no
ambiguity and no further freedom to make subtractions. Moreover, the
renormalization conditions are independent of the particular choice of
background $\sigma(\vek x)$, so it makes sense to compare results
for different choices of $\sigma(\vek x)$, {\it ie.\/} different
geometries.  Having been fixed in perturbation theory, the
counterterms are fixed once and for all and must serve to remove the
divergences that arise for any physically sensible $\sigma(\vek x)$. 
The Yukawa theory with coupling $g$ in three space dimensions gives a
textbook example: The $g\bar\psi\sigma\psi$ coupling generates
divergences in low order Feynman diagrams proportional to
$\sigma^{2}$, $\sigma^{4}$ and $(\partial\sigma)^2$ and therefore
requires one to introduce a mass, a quartic self-coupling, and a
kinetic term for $\sigma$.  This is the only context in which one can
study the fluctuations of a fermion coupled to a scalar background in
three dimensions.

In this Letter we study the vacuum fluctuations of a real scalar field
$\phi$ coupled to a scalar background $\sigma(\vek x)$ with coupling
$\lambda$, $\mathcal{L}_{\rm int}(\phi,\sigma)=\frac{1}{2}\lambda\,
\sigma(\vek{x})\,\phi^{2}(\vek{x},t)$.  In the limit where
$\sigma(\vek x)$ becomes a $\delta$-function on some surface ${\cal
S}$ and where $\lambda\to\infty$, it is easy to verify that all modes
of $\phi$ must vanish on ${\cal S}$.  We call this the \emph{Dirichlet
limit}.  It consists of the \emph{sharp limit}, where
$\sigma(\vek{x})$ gets concentrated on ${\cal S}$, followed by the
\emph{strong coupling limit}, $\lambda\to \infty$.  In general, we
find that the divergence of the vacuum energy in the Dirichlet limit
cannot be renormalized.  Generally, even the sharp limit does
not lead to a finite Casimir energy except in one dimension, where the
sharp limit exists but the Casimir energy diverges as
$\lambda\ln \lambda$ in the strong coupling limit.

This divergence indicates that \emph{the Casimir energy of a
scalar field forced to vanish on a surface in any dimension is
infinite}.  However, all is not lost.  The unrenormalizable
divergences are localized on ${\cal S}$, so quantities that do not
probe ${\cal S}$ are well defined.  For example, it is straightforward
to show that the vacuum \emph{energy density} away from ${\cal S}$ is
well defined in the Dirichlet limit, even though the energy
density on ${\cal S}$ diverges\cite{GR}.  We expect that this is true
in general.  The forces between rigid bodies are also finite in the
Dirichlet limit.  But any quantity whose definition requires a
deformation or change in area of ${\cal S}$ will pick up an infinite
contribution from the surface energy density and therefore diverge. 
For example, we will see explicitly that the vacuum contribution to
the \emph{stress} on a (generalized) Dirichlet sphere in two or
more dimensions is infinite.

The remainder of this Letter is organized as follows: First we briefly
review our computational method and discuss the structure of the
counterterms required by renormalization.  Then we present two
examples, leaving the details to Ref.~\cite{dens}.  We begin with the
simplest Casimir problem: two Dirichlet points on a line, where we can
compare our results with standard calculations that assume
boundary conditions from the start \cite{MT}.   We find that the
renormalized Casimir energy is infinite but the Casimir force is
finite in the Dirichlet limit.  We show how the QFT approach resolves
inconsistencies in the standard calculation.  Next we study the
Dirichlet circle in two dimensions.  We demonstrate explicitly that
the renormalized Casimir stress on the circle diverges in the
sharp limit. We show that the
divergence is associated with a simple Feynman diagram and will
persist in three dimensions (the ``Dirichlet sphere'') and beyond.

We define the bare Casimir energy to be the vacuum energy of a quantum
field $\phi$ coupled to a background field $\sigma$ by ${\cal L}_{\rm
int}(\phi,\sigma)$, minus the vacuum energy in the absence of
$\sigma$.  This quantity can be written as the sum over the
shift in the zero-point energies of all the modes of
$\phi$ relative to the trivial background
$\sigma=0$, $ E_{\rm bare}[\sigma]=\frac{\hbar}{2}\sum_n (\omega_n[\sigma]
- \omega_n^{(0)})$. Equivalently, using the effective action formalism of
QFT, $E_{\rm bare}[\sigma]$ is given by the sum of all $1$-loop
Feynman diagrams with at least one external $\sigma$ field.  The
entire Lagrangian is
\begin{equation}
    \mathcal{L} = \frac{1}{2} \partial_\mu\phi\,\partial^\mu\phi -
    \frac{m^2}{2} \phi^2 - \frac{\lambda}{2} \phi^2 \sigma(\vek{x}) +
    \mathcal{L}_{\rm CT}[\sigma]\,,
    \label{2.2}
\end{equation}
where ${\cal L}_{\rm CT}[\sigma]$ is the counterterm Lagrangian
required by renormalization.  Combining its contribution to the energy
with $E_{\rm bare}[\sigma]$ yields the renormalized energy $E_{\rm
cas}[\sigma]$.  We have taken the dynamics of the background field
$\sigma(\vek{x})$ to include only the $\phi$-$\sigma$ coupling and the
counterterms required by renormalization theory.  The frequencies
$\{\omega[\sigma]\}$ are determined by
$-\mathbf{\nabla}^{2}\phi(\vek{x}) + (m^2 +
\lambda\sigma(\vek{x}))\phi(\vek{x}) =
\omega^{2}[\sigma]\phi(\vek{x}).$ This is a renormalizable quantum
field theory, so $E_{\rm cas}[\sigma]$ will be finite for any smooth
$\sigma$ and finite $\lambda$.  We use the method developed in
Ref.~\cite{dens} to compute the Casimir energy of the background
configuration \emph{exactly} while still performing all the necessary
renormalizations in the perturbative sector.  The interested reader
should consult Ref.~\cite{GJW} for an introduction to the method and
Ref.~\cite{us} for applications.  We assume that the background field
$\sigma(\vek x)$ is sufficiently symmetric to allow the scattering
amplitude to be expanded in partial waves, which we label by $\ell$. 
We express the \emph{renormalized} Casimir energy as a sum over bound
states $\omega_{\ell j}$ plus an integral over continuum modes with $\omega =
\sqrt{k^2+ m^2}$,
\begin{widetext}\vspace*{-\bigskipamount}
\begin{equation}
E_{\rm cas}[\sigma]=\sum_\ell N_\ell\left[\sum_j
\frac{\omega_{\ell j}}{2} + \int_0^\infty \frac{dk}{2\pi}
\omega(k)\frac{d}{dk}\left[\delta_\ell(k)\right]_{N}\right]
+ E_{\rm FD}^{N} + E_{\rm CT}
\label{evac1}
\end{equation}
\end{widetext}
where $N_\ell$ denotes the multiplicity, $\delta_{\ell}$ the
scattering phase shift and $\frac{1}{\pi} \frac{d\delta_{\ell}}{dk}$
the continuum density of states in the $\ell^{\rm th}$ partial wave. 
The subscript $N$ on $\delta_{\ell}$ indicates that the first $N$
terms in the Born expansion of $\delta_{\ell}$ have been subtracted. 
These subtractions are compensated exactly by the contribution of the
first $N$ Feynman diagrams, $E^{N}_{\rm FD}=\sum_{i=1}^{N}E_{\rm
FD}^{(i)}$.  In eq.~(\ref{evac1}) $E_{\rm CT}$ is the contribution of
the counterterm Lagrangian, ${\cal L}_{\rm CT}$.  Both $E^{N}_{\rm
FD}$ and $E_{\rm CT}$ depend on the ultraviolet cutoff $1/\epsilon$,
but $E_{\rm FD}^{N} + E_{\rm CT}$ remains finite as $\epsilon\to 0$. 
One can think of $\epsilon$ as the standard regulator of dimensional
regularization, although our methods are not wedded to any particular
regularization scheme.  After subtraction, the $k$-integration in
eq.~(\ref{evac1}) converges and can be performed numerically for any
choice of $\sigma(\vek x)$.  It is convenient for computations to
rotate the integration contour to the imaginary $k$-axis giving
\begin{equation}
E_{\rm cas}[\sigma]=\sum\limits_{\ell} N_\ell \int\limits_m^\infty
\frac{d t}{2\pi}\,\frac{t}{\sqrt{t^2-m^2}}\,
\left[\beta_\ell(t)\right]_{N}
+ E_{\rm FD}^{N} + E_{\rm CT} \,,
\label{master}
\end{equation}
where $t=-ik$.  The real function $\beta_\ell(t)$ is
the logarithm of the \emph{Jost function} for imaginary momenta,
$\beta_\ell(t) \equiv \ln F_\ell(it)$.  Efficient methods to compute
$\beta_\ell(t)$ and its Born series can be found in Ref.~\cite{dens}. 
The renormalized Casimir energy density for finite $\lambda$,
$\epsilon_{\rm cas}(\vek x)$, can also be written as a Born subtracted
integral along the imaginary $k$-axis plus contributions from
counterterms and low order Feynman diagrams~\cite{dens}.

In less than three dimensions only the lowest order Feynman diagram
diverges, so only a counterterm linear in $\sigma$ is necessary,
$\mathcal{L}_{\rm CT} = c_1\lambda\sigma(\vek{x})$.  Since the tadpole
graph is also local, we can fix the coefficient $c_1$ by requiring a
complete cancellation, $E_{\rm FD}^{(1)} + E_{\rm CT} = 0$.  In three
dimensions it is necessary to subtract two terms in the Born expansion
of $\beta_{\ell}(t)$ and add back the two lowest order Feynman graphs
explicitly.  The counterterm Lagrangian must be expanded to include a
term proportional to $\sigma^{2}$, ${\cal L}_{\rm CT} =
c_1\lambda\sigma(\vek{x}) + c_2\frac{\lambda^2}{2}\,
\sigma^2(\vek{x})$.  The new term cancels the divergence generated by
the vacuum polarization diagram $E_{\rm FD}^{(2)}$, but it does not
completely cancel $E_{\rm FD}^{(2)}$, because $E_{\rm FD}^{(2)}$ it is
not simply proportional to $\int d^{3}x \sigma^{2}(\vek x)$.  To fix
$c_2$ we can only demand that it cancels $E_{\rm FD}^{(2)}$ at a
specified momentum scale $p^2 = M^2$.  Different choices of $M$
correspond to different models for the self-interactions of $\sigma$
and give rise to finite changes in the Casimir energy.  

We use eq.~(\ref{master}) and the analogous expression for the energy
density with backgrounds that are strongly localized about ${\cal S}$,
but not singular, to see how the Dirichlet limit is approached. 
It is straightforward to relate the Casimir energy density at the
point $\vek x$ to the Green's function at $\vek x$ in the background
$\sigma$, and then to show that it is finite as long as $\sigma(\vek
x)=0$.  Thus we find that the Casimir \emph{energy density} at any
point away from ${\cal S}$ goes to a finite limit as
$\sigma\to\delta_{\cal S}(\vek x)$ and $\lambda\to\infty$ and that the
result coincides with that found in boundary condition calculations. 
We also find a finite and unambiguous expression for the renormalized
Casimir energy density where $\sigma(\vek{x})$ is nonzero, as long as
it is nonsingular and the coupling strength is finite.  But as we
approach the sharp limit, the renormalized energy density on
${\cal S}$ diverges, and this divergence cannot be renormalized.

By analyzing the Feynman diagrams that contribute to the effective
energy we can deduce some general results about possible divergences
in the Casimir energy and energy density in the sharp limit.  In
particular, the divergences that occur in the Casimir energy in the
sharp limit come from low-order Feynman diagrams.  Specifically, using
dimensional analysis it is possible to show that in $n$ space
dimensions the Feynman diagram with $m$ external insertions of
$\sigma$ is finite in the sharp limit if $m>n$. 

Although less sophisticated methods can be used to obtain the energy
density at points away from ${\cal S}$ where renormalization is
unnecessary, as far as we know only our method can be used to define
and study the Casimir energy density where $\sigma(\vek{x})$ is nonzero
and therefore on ${\cal S}$ in the Dirichlet limit.   

Consider, as a pedagogical example, a real, massive scalar field
$\phi(t,x)$ in one dimension, constrained to vanish at $x=-a$ and $a$. 
The standard approach, in which the boundary conditions are imposed
{\it a priori\/}, gives an energy \cite{MT}
\begin{equation}
\widetilde{E}_{2}(a) = -\frac{m}{2} -
\frac{2a}{\pi}\int_{m}^{\infty}dt\frac{\sqrt{t^2-m^2}}{e^{4at}-1} \,,
\label{wrongenergy}
\end{equation}
where the tilde denotes the imposition of the Dirichlet boundary
condition at the outset.  From this expression one obtains an
attractive force between the two Dirichlet points, given by
\begin{equation}
\widetilde{F}(a)=-\frac{d\widetilde{E}_{2}}{d(2a)}
=-\int_{m}^{\infty}\frac{dt}{\pi}\frac{t^2}{\sqrt{t^2- m^2}(e^{4at}-1)}\, .
\label{force1}
\end{equation}
In the massless limit, we have $\widetilde{E}_{2}(a) = - \pi/48a$ and
$\widetilde{F}(a) = - \pi/96a^2$.

These results are not internally consistent, suggesting that the 
calculation has been oversimplified:  As $a\to\infty$,
$\widetilde{E}_{2}(a)\to -m/2$, indicating that the energy of an isolated
``Dirichlet point'' is $-m/4$.  As $a\to 0$ we also have a single
Dirichlet point, but $\widetilde{E}_{2}(a)\to\infty$ as $a\to 0$. 
Also note that $\widetilde{E}_{2}(a)$ is well defined as $m\to 0$, but
we know on general grounds that scalar field theory becomes infrared
divergent in one dimension when $m\to 0$.

We study this problem by coupling $\phi(t,x)$ to the static background
field $\sigma(x)=\delta (x+a) + \delta(x-a)$ with coupling strength
$\lambda$ as in eq.~(\ref{2.2}).  An elementrary calculation gives the
renormalized Casimir energy for finite $\lambda$,
\begin{widetext}\vspace*{-\bigskipamount}
\begin{equation}
E_2(a,\lambda)=\int_m^{\infty}\frac{dt}{2\pi}\frac{1}
{\sqrt{t^2-m^2}}\left\{ t \ln \left[
1+\frac{\lambda}{t}+\frac{\lambda^2} {4t^2}(1-e^{-4at})\right] -
\lambda\right\}
\label{e2}
\end{equation}
\end{widetext}
The same method can be applied to an isolated point giving,
\begin{equation}
    E_{1}(\lambda) =  \int_{m}^{\infty}\frac{dt}{2\pi}
    \frac{t\ln\left[1+\frac{\lambda}{2t}\right]
    -\frac{\lambda}{2}}{\sqrt{t^{2}-m^{2}}}
\end{equation}
For any finite coupling $\lambda$, the inconsistencies noted in
$\widetilde{E}_{2}(a)$ do not afflict $E_{2}(a,\lambda)$: As
$a\to\infty$, $E_{2}(a,\lambda)\to 2E_{1}(\lambda)$, and as $a\to 0$,
$E_{2}(a,\lambda)\to E_{1}(2\lambda)$.  Also $E_{2}(a,\lambda)$
diverges logarithmically in the limit $m\to 0$ as it should.  The
force, obtained by differentiating eq.~(\ref{e2}) with respect to
$2a$, agrees with eq.~(\ref{force1}) in the limit $\lambda \to
\infty$.

Note, however, that $E_{2}(a,\lambda)$ \emph{diverges} like 
$\lambda\log\lambda$ as $\lambda\to\infty$.  Thus the renormalized 
Casimir energy in a sharp background diverges as the Dirichlet boundary 
condition is imposed, a physical effect which is missed if the 
boundary condition is applied at the outset.

The Casimir \emph{energy density} for $x\ne\pm a$ can be calculated
assuming Dirichlet boundary conditions from the start simply
by subtracting the density in the absence of boundaries without
encountering any further divergences \cite{MT},
\begin{widetext}\vspace*{-\bigskipamount}
\begin{eqnarray}
\epsilon_{2}(x,a) & = &
-\frac{m}{8a}- \int_{m}^{\infty}\frac{dt}{\pi}
\frac{\sqrt{t^2-m^2}}{e^{4at}
-1} -\frac{m^2}{4a}\sum_{n=1}^{\infty}\frac{\cos
\left[\frac{n\pi}{a}(x-a)\right]}{\sqrt{(\frac{n\pi}{2a})^2+m^2}}
\mbox{ for } |x|<a \nonumber \\
\epsilon_{2}(x,a) & = & -\frac{m^2}{2\pi}K_0(2m|x-a|) \mbox{  for  } 
|x|>a \,.
\label{edens1}
\end{eqnarray}
\end{widetext}
The Casimir energy density for finite $\lambda$ was computed in
Ref.~\cite{dens}.  In the limit $\lambda\to\infty$ it agrees with
eq.~(\ref{edens1}) except at $x=\pm a$ where it contains an extra,
singular contribution.  If one integrates eq.~(\ref{edens1}) over all
$x$, ignoring the singularities at $x=\pm a$, one obtains
eq.~(\ref{wrongenergy}).  Including the contributions at $\pm a$ gives
eq.~(\ref{e2}).

This simple example illustrates our principal results: In the
Dirichlet limit the renormalized Casimir energy diverges because the
energy density on the ``surface,'' $x =\pm a$ diverges.  However the
Casimir force and the Casimir energy density for all $x\neq\pm a$
remain finite and equal to the results obtained by imposing the
boundary conditions {\it a priori\/}, eqs.~(\ref{force1}) and
(\ref{edens1}).

\emph{A scalar field in two dimensions} constrained to vanish on a circle of
radius $a$ presents a more complex problem. We decompose the energy
density in a shell of width $dr$ at a radius $r$ into a sum over angular
momenta, $\epsilon(r)= \sum_{\ell=0}^{\infty}
\epsilon_{\ell}(r)$, where $\epsilon_{\ell}(r)$ can
be written as an integral over imaginary momentum $t=-ik$
of the partial wave Green's function at coincident
points $G_{\ell}(r,r;it)$ and its radial derivatives.  First
suppose we fix $\sigma(\vek x)=\delta(r-a)$ and consider $r\ne a$.  It
is easy to see that the difference $[G_{\ell}(r,r,it)]_0$ 
between the full Green's function $G_{\ell}(r,r,it)$ and the free
Green's function $G_{\ell}^{(0)}(r,r,it)$ vanishes \emph{exponentially}
as $t\to\infty$.  For finite $\lambda$, both the $t$-integral and the
$\ell$-sum are uniformly  convergent so $\lambda\to\infty$ can be
taken under the sum and integral.  The resulting energy density,
$\tilde\epsilon(r)$, agrees with that obtained when the Dirichlet
boundary condition, $\phi(a)=0$, is assumed from the start.  As in one
dimension, nothing can be said about the total energy because
$\tilde\epsilon(r)$ is not defined at $r=a$, but unlike that case  the
integral of $\tilde\epsilon(r)$ now diverges even in the sharp limit for
finite $\lambda$.

To understand the situation better, we take $\sigma(\vek x)$ to be a
narrow Gau{\ss}ian of width $w$ centered at $r=a$ and explore the
sharp limit where $w\to 0$ and $\sigma(\vek
x)\to\delta(r-a)$.  For $w\ne 0$, $\sigma$ does not vanish at any
value of $r$, so $[G_{\ell}(r,r,it)]_0$ no longer falls exponentially
at large $t$, and subtraction of the first Born approximation to
$G_{\ell}(r,r,it)$ is necessary.  As in one dimension, the
compensating tadpole graph can be canceled against the counterterm,
$c_{1}\lambda\sigma(\vek x)$.  The result is a renormalized Casimir
energy density, $\epsilon(r,w,\lambda)$, and Casimir energy,
$E(w,\lambda)=\int_{0}^{\infty}dr \epsilon(r,w,\lambda)$, both of
which are finite.  However as $w\to 0$ both $\epsilon(a,w,\lambda)$
and $E(w,\lambda)$ diverge, indicating that the renormalized Casimir
energy of the Dirichlet circle is infinite.

The divergence originates in the order $\lambda^2$
Feynman diagram.  We study this diagram by subtracting the
\emph{second} Born approximation to $G_{\ell}(r,r,it)$ and adding back
the equivalent diagram explicitly.  Then the  $\ell$-sum and
$t$-integral no longer diverge in the  sharp limit.  In
the limit $w\to 0$, the diagram contributes
\begin{equation}
\lim_{w\to 0}E_{\rm FD}^{(2)} = - \frac{\lambda^2\,a^2}{8}\,
\int_0^\Lambda dp\,J_0^2(a p)\,\arctan\frac{p}{2m}  
\label{B6}
\end{equation}
which diverges like $\ln\Lambda$.  The divergence originates in the
high momentum components in the Fourier transform of
$\sigma(r)=\delta(r-a)$, not in the high energy behavior of a
loop integral, and therefore cannot be renormalized.  Taking
$\lambda\to\infty$ only makes the divergence worse.  Because it varies
with the radius of the circle, this divergence gives an infinite
contribution to the surface tension.  This divergence only gets
worse in higher dimensions (in contrast to the claim of
Ref.~\cite{Milton}).  For example, for $\sigma(r)=\delta(r-a)$ in
three dimensions the \emph{renormalized} two point function is
proportional to an integral over $p$ of a function proportional to
$\lambda^{2}a^{4} p^{2} j_{0}^{2}(pa)\ln p$ at large $p$.  The
integral diverges like $\Lambda\ln\Lambda$. Such divergences cancel
when we compute the force between rigid bodies, but not in the case of
stresses on isolated surfaces.

\emph{In summary}, by implementing a boundary condition as the limit
of a less singular background, we are able to study the divergences
that arise when a quantum field is forced to vanish on a prescribed
surface.  For all cases we have studied, the renormalized
Casimir energy, defined in the usual sense of a continuum quantum
field theory, diverges in the Dirichlet limit. Physical cutoffs (like
the plasma frequency in a conductor) regulate these divergences,  
which are localized on the surface.  On the other hand the
energy density away from the surfaces or quantities like the force
between rigid bodies, for which the surfaces can be held fixed, are
finite and independent of the cutoffs.  Observables that require a
deformation or change in area of ${\cal S}$ cannot be defined
independently of the other material stresses that characterize the
system.  Similar studies are underway for fluctuating fermion
and gauge fields, leading to Neumann and mixed boundary conditions
with the same types of divergences.  \bigskip

\paragraph{Acknowledgments}
We gratefully acknowledge discussions with G.~Barton, E.~Farhi and
K.~D.~Olum.  N.~G. and R.~L.~J. are supported in part by the
U.S.~Department of Energy (D.O.E.) under cooperative research
agreements~\#DE-FG03-91ER40662 and~\#DF-FC02-94ER40818.  M.~Q. and
H.~W. are supported by the Deutsche Forschungsgemeinschaft under
contracts~Qu 137/1-1 and~We~1254/3-2.

%%%%%%%%%%%%%%%%%%%%%%%%%%%%%%%%%%%%%%%%%%%%%%%%%%%%%%%%%%%%%%%%%%%%%%%%%%%

%%%%

\end{document}